\documentclass[10pt,a4paper,final]{iopart}

\usepackage{iopams}  
\usepackage{graphicx}
\usepackage[breaklinks=true,colorlinks=true,linkcolor=blue,urlcolor=blue,citecolor=blue]{hyperref}
\usepackage{changes}
\colorlet{Changes@Color}{red}

\begin{document}

\title[Sn ion energy distributions from laser produced plasmas]{Sn ion energy distributions of ns- and ps-laser produced plasmas}

\author{A. Bayerle$^{1}$, M.~J. Deuzeman$^{1,2}$, S. van der Heijden$^{1}$, D. Kurilovich$^{1,3}$, T. de Faria Pinto$^{1}$, A. Stodolna$^{1}$, S. Witte$^{1,3}$, K.~S.~E. Eikema$^{1,3}$, W. Ubachs$^{1,3}$, R. Hoekstra$^{1,2}$, and O.~O. Versolato$^{1}$}
\address{$^1$Advanced Research Center for Nanolithography (ARCNL), \mbox{Science Park 110},  \mbox{1098 XG} Amsterdam, The Netherlands}
\address{$^2$Zernike Institute for Advanced Materials, University of Groningen, \mbox{Nijenborgh 4}, \mbox{9747 AG} Groningen, The Netherlands}
\address{$^3$Department of Physics and Astronomy and LaserLab, Vrije Universiteit Amsterdam, \mbox{De Boelelaan 1081}, \mbox{1081 HV} Amsterdam, The Netherlands}
\ead{o.versolato@arcnl.nl}

\begin{abstract}
Ion energy distributions arising from laser-produced plasmas of Sn are measured over a wide laser parameter space. Planar-solid as well as liquid-droplet targets are exposed to infrared laser pulses with energy densities between 1\,J/cm$^2$ and 4\,kJ/cm$^2$ and durations spanning 0.5\,ps to 6\,ns. 
The  measured ion energy distributions are compared to two self-similar solutions of a hydrodynamic approach assuming isothermal expansion of the plasma plume into vacuum. 
For planar and droplet targets exposed to ps-long pulses we find a good agreement between the experimental results and the self-similar solution of a semi-infinite simple planar plasma configuration with an exponential density profile. The ion energy distributions resulting from solid Sn exposed to ns-pulses agrees with solutions of a limited-mass model that assumes a Gaussian-shaped initial density profile.
\end{abstract}
\vspace{2pc}
\maketitle
\ioptwocol
\section{Introduction}
Plasma expansion into vacuum is a subject of great interest for many applications ranging from ultracold plasmas \cite{Cummimgz2005,Killian1999} over laser acceleration \cite{Flacco2008laseracceleration,Mckenna2006laseracceleration} to short-wavelength light sources \cite{BakshiSourcesBook,Banine2011}. For such light sources driven by laser-produced plasmas (LPPs) the optics that collect the plasma-generated light are exposed to particle emission from the plasma. The impinging particles may affect the performance of the light-collecting optics.\\\indent
Charged particles from LPPs can be monitored by means of Faraday cups (FCs) - a robust plasma diagnostics tool. Faraday cups can be used to characterize the angular distribution of ion emission of metal and non-metal LPPs \cite{thestrup2003,ToftmannfsAblationAg}. Faraday cups in time-of-flight mode can be used to measure the energy distributions of the ions emanating from the plasma interaction zone \cite{verhoff2012,Freeman2012,FaridIonsLPP2013}. Because of its relevance to extreme ultraviolet nanolithography, LPP of Sn has been subject to similar studies, in which the kinetic energy and yield of the Sn ions together with extreme-ultraviolet light output is characterized \cite{Coons2010}. Indications of a set of laser parameters was reported for which a dip in the Sn ion yield might occur \cite{ISANVinokhodovIonDip2014}. Both droplet and planar targets have been investigated \cite{chen_wang_zuo_wang_2016,Deuzeman2017} but no unique optimal conditions have been found so far. \\\indent In order to understand the ion energy distributions from LPPs, a theoretical framework based on hydrodynamic expansion has been established early on \cite{Gurevich1966,AnisimovGasDyn1993}. The theoretical framework has been expanded ever since. Nevertheless, benchmarking the energy distribution functions derived in the different studies with experimental data on LPPs remains scarce. To the best of our knowledge only two groups report the comparison of the results of hydrodynamics models to ion energy distributions measured by FCs \cite{doggett2011,Murakami2005}. 
\\\indent Laser-produced plasmas can be created over a vast space of laser and target parameters. Here we address the energy distributions of emitted ions in a substantial subset of this space, namely pulse lengths ranging from sub-ps to almost 10\,ns and laser peak fluences up to 3\,kJ/cm$^2$. 
The plasma is produced on solid-planar and liquid-droplet targets irradiated by infrared lasers. The measured results are used to benchmark two analytical solutions of hydrodynamics models of plasma expansion into vacuum \cite{Murakami2005,Mora2003PlasmaExpansionVacuum}. \added{The intended accuracy of this comparison between theory and our experiments is not expected to be able to discern any effects beyond those predicted by these single-fluid single-temperature hydrodynamic plasma models, such as the possible presence of a double layer \cite{Bobrenok2000,Mora2003PlasmaExpansionVacuum,Murakami2006}.} First, the solution to a semi-infinite simple planar model assuming an exponential density profile of the plasma \cite{Mora2003PlasmaExpansionVacuum} shows good agreement with the experimental results of LPP by ps-laser pulses. Second, the ion energy distributions obtained by exposing solid Sn targets to 6-ns laser pulses agrees best with the solution to a modified hydrodynamics model \cite{Murakami2005}. In that work, a different density evolution of the expanding plasma is derived, starting out from a Gaussian density profile instead of the exponential profile used in the work of Mora \cite{Mora2003PlasmaExpansionVacuum}. 
In addition, the modified model takes into account the dimensionality of the plasma expansion. 
\\\indent In Sec.~\ref{sec:experiment} the experimental setups used to produce Sn plasmas by pulsed lasers are described. The ion energy distributions are shown in Sec.~\ref{sec:results}. 
We compare the ion energy distributions with the results of theoretical studies on plasma expansion into vacuum which are briefly reviewed in the following Sec.~\ref{sec:theorie}.
\section{Theoretical models}
\label{sec:theorie}
Plasma expansion into vacuum traditionally is treated by a hydrodynamic approach \cite{Gurevich1966}. \deleted{In most cases t} \added{A typical} initial condition consists of cold ions with a charge state $Z$ and a \added{hot} gas of electrons with energies distributed according to Maxwell-Boltzmann \cite{crow_auer_allen_1975}. 
The electron cloud overtakes the ions during expansion leading to an electrostatic potential that accelerates the ions. The hydrodynamic equations \added{of plasma expansion} can be solved by a self-similar ansatz with the coordinate $x/R(t)$, where $x$ is is the spatial coordinate and $R(t)=c_{\rm s} t$ \cite{Mora2003PlasmaExpansionVacuum} or $R(t) \propto t^{1.2}$ \cite{Murakami2005} is the characteristic system size growing with the sound speed $c_{\rm s}$. Many theoretical studies that are based on such a hydrodynamics approach solve the problem of plasma expansion into vacuum by making different assumptions, for example 
isothermal or adiabatic expansion \cite{SACK1987311} or a non-Maxwellian distribution of the electrons \cite{Alexey2009,bennaceur-doumaz_bara_djebli_2015}. Here we focus on two studies published by Mora \cite{Mora2003PlasmaExpansionVacuum} and Murakami \textit{et al.} \cite{Murakami2005} \added{where we assume that the charge state $Z$ can be interpreted as an average charge state. This presents a strong simplification especially in our rapidly expanding laser-driven plasma containing multiply charged ions (e.g., see  Refs.~\cite{Fujioka2005,AndreaThumJager1999}).  Our FC technique cannot resolve ions by their charge and the measured distribution is in fact a convolution of distributions of ions of the various charge states. These energy distributions may be expected to depend on charge state $Z$ (see, e.g., Refs.~\cite{Bobrenok2000,apinaniz2015})
and the collected charge on the FC is $Z$ times the amount of ions captured.”} 
\deleted{plasma and provide solutions to the models in terms of emitted particle number per energy interval.} 
\deleted{Although laser-produced plasmas contain many possible charge states of the target element} \added{Nevertheless, } it is instructive to compare the charge-per-ion energy distributions measured on FCs with the \added{solutions to these single-fluid single-temperature hydrodynamic plasma models in terms of emitted particle number per energy interval.} \deleted{following distributions.} In Mora \cite{Mora2003PlasmaExpansionVacuum} the particle energy distribution is found to be 
\begin{equation}
dN/dE \propto \left(E/E_{0}\right)^{-1/2} \exp\left(-\sqrt{E/E_{0}}\right),
\label{MorasModelEq}
\end{equation}
while Murakami \textit{et al.} \cite{Murakami2005} derives
\begin{equation}
dN/dE \propto \left(E/\tilde{E_{0}}\right)^{(\alpha-2)/2} \exp\left(-E/\tilde{E_{0}}\right),
\label{MurakamiModelEq}
\end{equation}
under inclusion of higher dimensionality $\alpha$ and Gaussian evolution of the density. \\\indent The respective ion energies are characterized by $E_0$ or $\tilde{E_0}$. The characteristic energy dependents on the charge state $Z$ of the ions and the electron temperature $T_e$. In the first equation the characteristic ion energy $E_0$ is given by
\begin{equation}
E_{0} = Z k_{\rm B}T_e,
\label{eq:eoformula}
\end{equation}
with $k_{\rm B}$ the Boltzmann constant. The ion energy in Eq.~(\ref{MurakamiModelEq}) is given by
\begin{equation}
\tilde{E_{0}} = m \dot{R}^2(t)/2 =  2 Z k_{\rm B} T_e \ln{\left(R(t)/R_0\right)} ,
\label{eq:eoformula2}
\end{equation}
with $m$ the ion mass and $R_0$ the initial size.
A higher $E_{0}$ or $\tilde{E_0}$ mean there are relatively more high-energy ions, with a higher mean charge state and a higher electron temperature.
\\\indent Both models assume Boltzmann-distributed electron energies and isothermal expansion of the plasma. \added{Additionally, in Ref.~\cite{Murakami2005} the solution (our Eq.\,(\ref{MurakamiModelEq})) is extended and smoothly connected with a solution of an adiabatically expanding plasma. The resultant ion energy spectrum is given in the same form as our Eq.\,(\ref{MurakamiModelEq}) only with a slight modification in the characteristic energy scale \mbox{$\tilde{E}_0 \rightarrow f \tilde{E}_0$}. For simplicity, we use the solution in their first step to analyze our experimental results.}
\\\indent One essential difference between the two models is the functional form of the density evolution of the expanding plasmas. In Ref.~\cite{Mora2003PlasmaExpansionVacuum}, the charge density is obtained as a perturbation of the initial charge density, which then evolves as $n \propto \exp{(-x/R(t))}$ (see also Ref.~\cite{LandauLifshitzFM}). In Ref.~\cite{Murakami2005} the authors argue that for longer pulse lengths or limited target masses this perturbation assumption is not valid. They obtain a Gaussian form for the charge density profile \cite{London1981,Robicheaux2003}: $n \propto \exp{(-(x/R(t))^2)}$. This density profile results in a different high-energy tail of the ion distribution. The dimensionality is captured by the parameter $\alpha$. If $\alpha = 1$, the expansion is planar otherwise the expansion is cylindrical or spherical for $\alpha = 2$ and $\alpha = 3$ respectively.
\section{Experimental setup}
\label{sec:experiment}
\begin{figure}[tb!]
	\centering
	\includegraphics{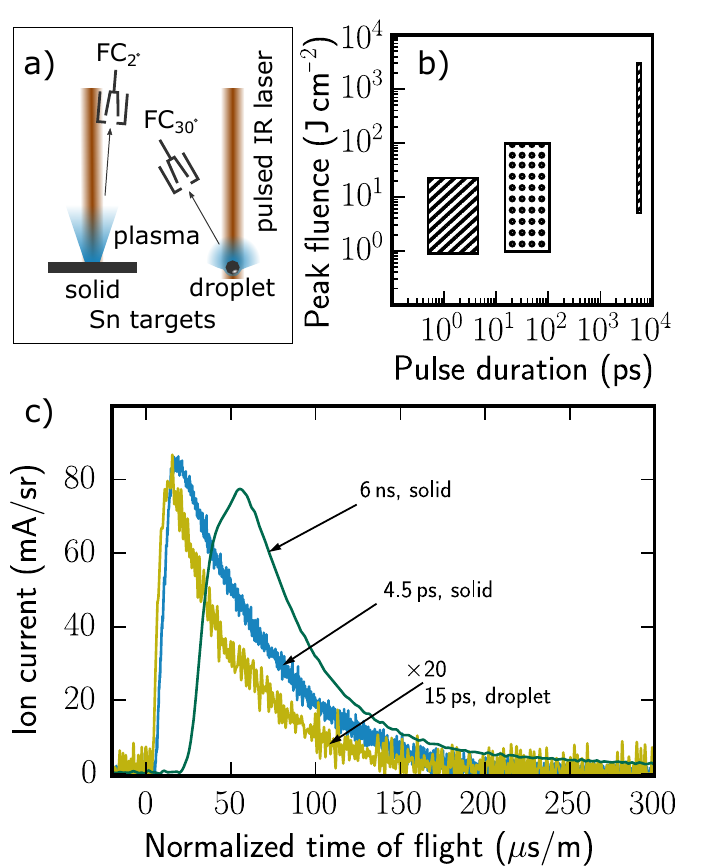}
	\caption{\label{fig:setupandTOFs}a) Schematic of the experimental setups. The plasma is created by exposing Sn metal targets to focused infra-red laser pulses. The Sn target has either planar geometry (solid target) or consists of droplets of 30\,$\mu$m diameter. The ion emission is collected by Faraday cups (FC) that are roughly 1\,m away from the plasma source. b) Pulse duration and peak fluence parameter space addressed by the experiments. Hatched rectangles show the parameter space explored using solid targets. The parameter space explored on Sn droplets is shown by the dotted rectangle. c) Typical examples of time dependent ion traces collected by the FCs. The x-axis is normalized to a time-of-flight distance of 1\,m. The targets are exposed to fluences of 25\,J/cm$^2$ (solid target) and 30\,J/cm$^2$ (droplet target)}
\end{figure} 
We use two setups to create laser-produced plasmas of Sn and measure the energy distributions of the emitted ions. Figure~\ref{fig:setupandTOFs}a. shows the schematic representation of the setups. The first setup contains a solid Sn plate of 1\,mm thickness as a target. In the second experiment the targets are free falling droplets of molten Sn with a diameter of 30\,$\mu$m. The solid and droplet targets reside in vacuum apparatuses with base pressures below 10$^{-6}$\,mbar. Pulsed infrared laser beams are focused on the targets to create the plasma. The ion emission is collected by FCs mounted into the vacuum apparatus around the plasma.
\\\indent
The custom-made FCs consist of a cone shaped charge collecting electrode mounted behind a suppressor electrode \cite{Deuzeman2017}. Both electrodes are housed in a grounding shield. The FCs have an opening of 6\,mm diameter and are mounted at a distances between 25\,cm and 75\,cm.
The collector and suppressor are biased to a negative potential with respect to ground in order to prevent plasma electrons from entering the cup, and secondary electrons from leaving the cup after Sn ions impinge on the surface of the collector. \\\indent Faraday cup measurements can only serve to give an approximation of the plasma flow as the separation of electrons from the ions in the quasi-neutral expansion of the plasma cannot be assumed to be complete and may depend on the set bias voltages and earth magnetic fields \cite{PELAH1976348}. We verified that further increasing the bias voltages had no significant impact on the measured time-of-flight traces. The earth magnetic field is only expected to influence the detection of low-energy ions. 
\\\indent
Figure~\ref{fig:setupandTOFs}c shows typical time-of-flight traces acquired by the FCs during experimental runs. The ion current is measured across a shunt resistor with a digital storage oscilloscope. The traces are averaged for the same laser fluence for about hundred laser exposures. The ns-laser produced traces have a lower noise amplitude, because the traces are averaged for about two hundred exposures. \added{The} shunt resistor \added{of 10\,k$\Omega$} and the added capacitance of \added{220\,pF} of the collector cup and the cable to the oscilloscope form an RC-network that limits the bandwidth of the measurement. The effective RC-time of the read-out is on the order of 2\,$\mu$s. In order to retrieve the ion current from the raw data we correct for the response function of the read-out network. The ion traces can be integrated in time to obtain the total charge emitted into the direction of the corresponding FC. The energy distribution can be calculated by the following transformation
\begin{equation*}
dQ/dE = t^3 I(E) / m L, \; E = m L^2 / 2 t^2,
\end{equation*}
with $m$ the mass of Sn, $L$ the distance between the plasma and the detector and $t$ the time-of-flight. The charge yield per energy interval is averaged over bins of 10\,eV. \\\indent As shown in Fig.~\ref{fig:setupandTOFs}c, the time-of-flight traces for pulses below 15\,ps have a smaller signal-to-noise ratio. The traces converge to the background noise level at 170\,$\mu$s/m. This time-of-flight is equivalent to an energy of 20\,eV. Therefore we truncate the energy distributions below 20\,eV.
\\\indent The setup containing the droplet target is described in detail by Kurilovich \textit{et al.} \cite{Kurilovich2016}. The Sn droplets are created by pushing liquid Sn through a piezo-driven orifice. Orifice diameter and piezo driver frequency determine the diameter of the droplets to 30\,$\mu$m. A pulsed 1064-nm Nd:YAG laser is focused to a 100\,$\mu$m full width at half maximum (FWHM) Gaussian spot at the position of the droplet stream. Faraday cups are added at 37\,cm under angles of 30$^{\circ}$ and 60$^{\circ}$ with respect to the incoming laser beam to enable time-of-flight measurements.\\\indent
The second setup containing the solid target is described in detail by Deuzeman et al. \cite{Deuzeman2017}. The solid target is mounted onto a 2D-translation stage (PI miCos model E871) enabling a computer-controlled, stepwise motion of the target between laser pulses in perpendicular direction to the laser beam. The stepwise translation of the target between pulses is necessary to prevent the ion emission to change because of surface deformation after too many laser shots on the same spot. Also, the first few laser pulses on a new spot on the surface ablate the oxide layer and the subsequent laser pulses produce plasmas containing mostly Sn \cite{BuchBaaerle}. Two laser systems are employed to create plasma at the Sn solid surface. First, a 800-nm wavelength Ti:sapphire laser is used to generate pulses of 0.5\,ps to 4.5\,ps duration. The Gaussian spot size of the the 800-nm laser at the surface of the target is 100\,$\mu$m FWHM.
Second, a Nd:YAG laser outputs 6-ns long pulses. This laser has a wavelength of 1064\,nm and is focused to a Gaussian spot of 90\,$\mu$m FWHM. The setup is equipped with three FCs, one at a distance of 73\,cm and at an angle of 2$^{\circ}$ from the surface normal, and two at $\pm$30$^{\circ}$ at distances of 26\,cm and 73\,cm.
\\\indent We summarize the laser parameter space accessible with the lasers in Figure~\ref{fig:setupandTOFs}b. The peak fluence and pulse duration used in the experiments performed on a solid target are shown as hatched rectangles. The Ti:sapph laser produces ultrashort pulses ranging from 0.5\,ps to 4.5\,ps without evidence for intensity-induced self-focusing or self-phase modulation effects. Peak pulse energy densities run up to 30\,J/cm$^2$. The pulse length of the Nd:YAG laser used on the solid target is 6\,ns and the pulse energy densities reach 3\,kJ/cm$^2$. The dotted rectangles shows the parameter space for the experiments on droplets. The Nd:YAG laser employed in the droplet setup is capable of producing ultrashort pulses between 15\,ps and 105\,ps duration and peak fluences of 1 to 100\,J/cm$^2$
\section{Results and discussion}
\label{sec:results}
First we present the energy distributions of the Sn ion emission for three different pulse lengths and same energy density of the laser and show that the experimental data can be well described by the self-similar solutions of the hydrodynamic model. Second, we show the ion distributions obtained for different laser fluences and for fixed pulse durations.
\subsection{Changing pulse duration}
\begin{figure}[tb!]
	\centering
	\includegraphics{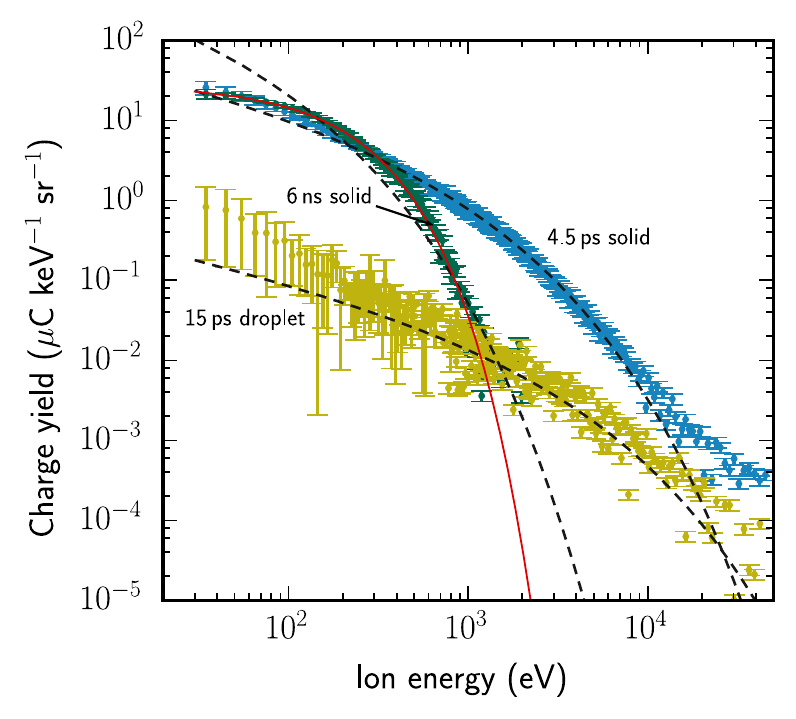}
	\caption{\label{fig:VarypulseLent} Charge energy distributions measured for different pulse durations of the laser on both solid-planar and liquid-droplet targets. The energy density of the laser pulses is $25-30$\,J/cm$^2$. The dashed (black) lines show the fits of Eq.~(\ref{MorasModelEq}) to the distributions. The solid (red) line is a fit of Eq.~(\ref{MurakamiModelEq}) with $\alpha = 2$ to the data.}
\end{figure}
We measure the ion energy distributions on the different target geometries with the following laser parameters. The solid target is irradiated by 6-ns, 1064-nm and 4.5-ps, 800-nm pulses with a peak fluence of 25\,J/cm$^2$ and the Sn droplets are exposed to 15-ps and 105-ps pulses with a peak fluence of 30\,J/cm$^2$ and 1064\,nm wavelength. The presented ion energy distributions are measured under different angles for the two target geometries. Ion emission from the solid target is measured at \mbox{$2^{\circ}$} \added{(and \mbox{30$^{\circ}$}, see Deuzeman et al. \cite{Deuzeman2017})} with respect to the surface normal, while the droplet target emission is collected by the FC mounted at an angle of \mbox{30$^{\circ}$} from the laser axis. \added{Because most (and most energetic) ions are emitted along the surface normal \cite{AndreaThumJager1999,AndreaThumJager2000,ZiqiChen2016} the ion emission in the \mbox{30$^{\circ}$} direction from the spherical droplet target (thus emitted along a surface normal) is best compared to the ion emission in the small-angle, \mbox{2$^{\circ}$} direction from the planar target. In this comparison we note that the projection of the laser beam onto the droplet surface at a \mbox{30$^{\circ}$} angle-of-incidence will reduce the local fluence by the cosine of this angle. The absorption, governed by the Fresnel equations, also depends on this angle. Both effects, however, have minor impact considering the relatively small angle involved and, in fact, these two effects partially cancel each other (see, e.g., Ref.~\cite{Basko201759}). The difference in the reflectivities between solid and liquid tin before laser impact is quite small at 2 percentage points, comparing 82 to 84\%, respectively (taking as input the works of Refs.~\cite{Cisneros1982,Golovashkin1966}. At our typical energy fluences, however, the solid target is practically instantaneously melted and heated to several thousand degrees (within the skin layer). Thus, the target reflectivity, identically for both solid planar and liquid droplet cases, is determined by the optical properties of liquid and vaporized tin at  T $\sim$ 3000\,K-5000\,K that are poorly known and quite different from those at room temperature. }\\\indent
Figure~\ref{fig:VarypulseLent} shows the ion energy distributions of the LPPs obtained with the laser parameters described above. In all cases the charge yields decrease monotonically with ion energy. Charge yields obtained from pulses below 6-ns duration converge and hit the detection threshold around an ion energy of 30\,keV. Long laser pulses of 6\,ns produce charge yields that roll off already at 1\,keV at a faster rate. \\\indent
For ps-pulses the charge yield retrieved from the solid target is more than an order of magnitude higher than from the droplet target for energies below 5\,keV. 
For the solid target we acquire a total charge of about 4\,$\mu$C/sr and 3\,$\mu$C/sr for 4.5-ps and 6-ns pulse length, respectively. The droplet target yields a total charge of only 0.06\,$\mu$C/sr when exposed to the 15-ps laser pulse. We attribute this difference between collected charge to the smaller droplet diameter compared to the focused laser beam diameter. 
While the solid target is irradiated by a full Gaussian intensity profile, the droplet is exposed to only a fraction of the focused laser beam energy because the diameter of the droplet is three times smaller than the FWHM of the beam. The energy deposited on the droplet can be calculated by integrating the Gaussian beam fluence profile over the droplet. Then the energy on the droplet is \mbox{$E_d = E_{\rm L} (1-2^{-d_{\rm D}^2/d_{\rm L}^2})$} with $d_{\rm D}$ the droplet diameter, $E_{\rm L}$ and $d_{\rm L}$ the total laser energy and the FWHM diameter of the focused laser beam. For our experimental parameters the droplet is exposed to only 6\% of the total laser energy and thus the observed total charge yield will be substantially smaller than from the solid target. 
\\\indent The energy distributions of Fig.~\ref{fig:VarypulseLent} are compared with the theoretical predictions discussed above. The dashed (black) lines show the least-squares fitted energy distributions according to Eq.~(\ref{MorasModelEq}) for pulse lengths of 4.5\,ps and 15\,ps. The experimental energy distributions agree well with Eq.~(\ref{MorasModelEq}) for both target geometries and slightly different wavelengths. Applying the model comparison yields the characteristic ion energy $E_{\rm 0}$. For the 4.5-ps LPP we obtain \mbox{$E_{\rm 0}=250(30)$\,eV}. \\\indent
Model comparisons of the energy distributions of Sn ions emitted from the droplet target give higher characteristic energies. The plasma produced by the 15-ps laser pulses with 30\,J/cm$^2$ energy density yields \mbox{$E_{\rm 0} = 970(120)$\,eV}. This higher characteristic energy could well be the result of the irradiation of the droplet by only the central fraction of the laser beam where the fluence is highest. The droplet is exposed to the central 6\% of the total laser energy, therefore the average fluence is close to the peak fluence and thus exceeds the one on the solid target. 
\\\indent Irradiating the solid target surface with the 6-ns laser pulses produces an energy distribution that does not agree with Eq.~(\ref{MorasModelEq}) as illustrated in Fig.~\ref{fig:VarypulseLent} by the dashed (black) line.
The fit of Eq.~(\ref{MurakamiModelEq}) to the measured energy distribution is shown as a solid (red) line in Fig.~\ref{fig:VarypulseLent}. The dimensionality parameter is set to $\alpha=2$ and with a characteristic ion energy of $\tilde{E_{\rm 0}}=150(15)$\,eV, the model agrees well with the measured distributions. 
\\\indent The energy distributions of LPP Sn ions are reproduced well in the energy interval of 20\,eV to 20\,keV, although the target geometries and pulse durations vary significantly. Laser produced plasmas of ps-pulses show good agreement with Eq.~(\ref{MorasModelEq}), and can thus be modelled by the approach of Mora \cite{Mora2003PlasmaExpansionVacuum}. Between 100\,ps and 6\,ns pulse duration the ablated target material starts to absorb the laser energy and the density profile deviates from \mbox{$\rho\propto \exp{(-x/R(t))}$}. In this case we cannot expect Eq.~(\ref{MorasModelEq}) to fit the data. Instead, the experimental energy distribution for the 6-ns laser produced plasma is well described by Eq.~(\ref{MurakamiModelEq}).
\\\indent In the following, we focus on the study of the applicability of the two introduced models over the measured range of laser energy densities.
\subsection{Changing laser energy density}
\begin{figure}[bt!]
	\centering
	\includegraphics{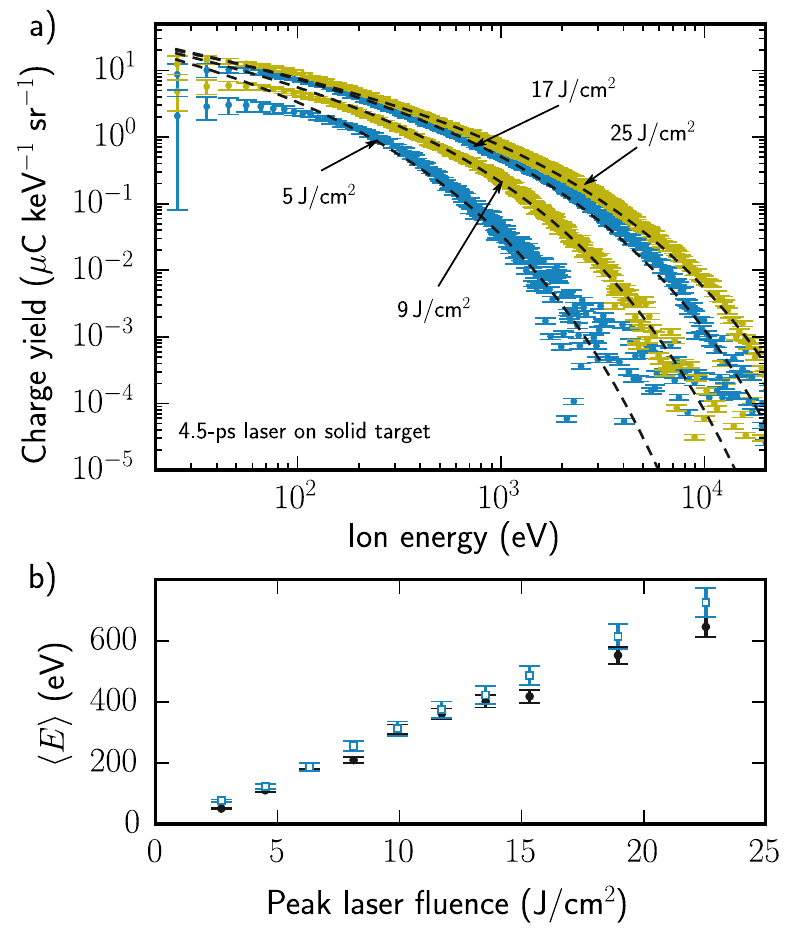}
	\caption{\label{fig:soligPS_fluence} (a) Charge energy distributions for a pulse duration of 4.5\,ps and different laser energy densities on the solid target, and fits with Eq.~(\ref{MorasModelEq}).
    (b) The values for $2 E_{\rm 0}=\langle E_{\rm fit} \rangle$ (solid, black circles) obtained from the fits with Eq.~(\ref{MorasModelEq}) for these distributions, with $\langle E_{\rm exp} \rangle$ (open, blue squares).}
\end{figure}
In the following we explore the applicability of the two models to ion energy distributions obtained from LPPs at different energy densities of the laser and fixed pulse durations.
\\\indent
The solid target is exposed to 4.5-ps pulses from the Ti:sapph laser with different energy densities. The resulting charge energy distributions are shown in Fig.~\ref{fig:soligPS_fluence}a. The four plots on the top are acquired by the FC at 2$^{\circ}$. These energy distributions are fit with Eq.~(\ref{MorasModelEq}) and shown as dashed (black) lines. 
\added{It is informative to compare also the average kinetic energies obtained from the fits $\langle E_{\rm fit} \rangle$ to those obtained directly from the data $\langle E_{\rm exp} \rangle$ enabling to judge how accurately the theories describe the experiments. The average energy $\langle E_{\rm fit} \rangle = 2E_0$ and $\langle E_{\rm fit} \rangle = \tilde{E}_0/2$ for $\alpha=1$ can be obtained from Eq.~(\ref{MorasModelEq},\ref{MurakamiModelEq}) analytically but a correction related to the low-energy, 20\,eV cut-off needs to be applied to the values $\langle E_{\rm exp} \rangle$. The corresponding correction factor ranging from 1.2 to 1.6 is obtained by comparing the energy averages of Eqs.\,(\ref{MorasModelEq},\ref{MurakamiModelEq}) from zero to infinity and from 20\,eV and infinity. The correction factor is applied to $\langle E_{\rm exp} \rangle$ in the following. We find good agreement between the obtained values as presented in Fig.\,\ref{fig:soligPS_fluence}b.}
\begin{figure}[bt!]
	\centering
	\includegraphics{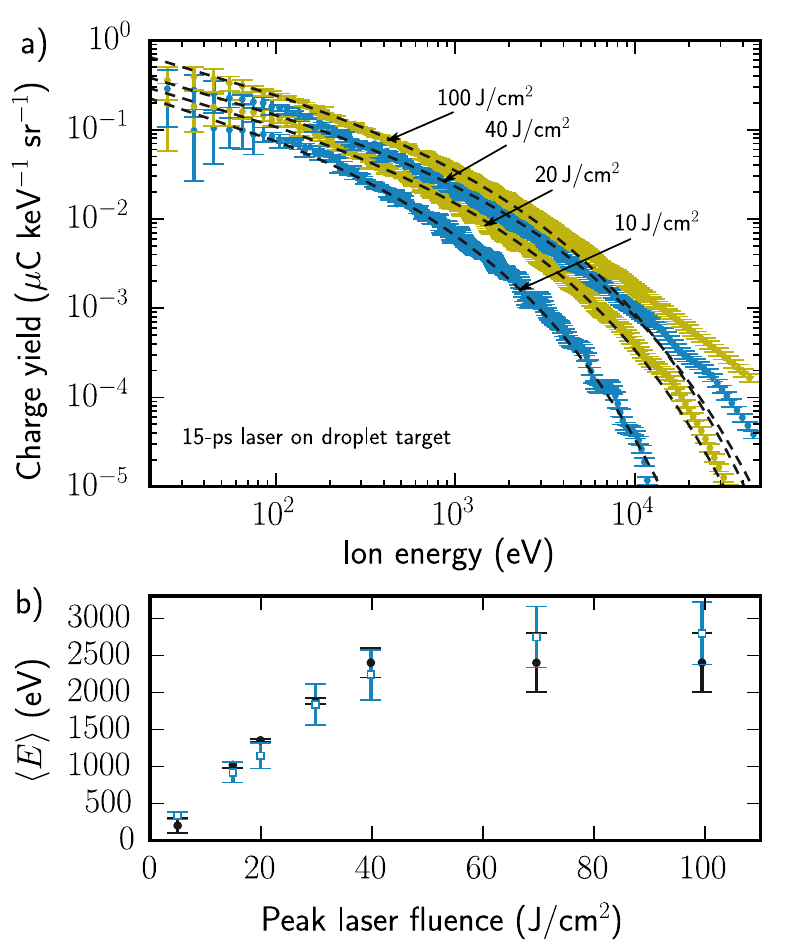}
	\caption{\label{fig:psDroplet} (a) Charge yield distributions for different energy densities of the laser on the Sn droplets and fits with Eq.~(\ref{MorasModelEq}). \deleted{The peak fluences are 10, 20, 40, 100\,J/cm$^{2}$ from bottom to top graph. The pulse duration of the laser pulses is 15\,ps.}(b) The values for $2 E_{\rm 0}=\langle E_{\rm fit} \rangle$ (solid, black circles) obtained from the fits with Eq.~(\ref{MorasModelEq}) for these distributions, with $\langle E_{\rm exp} \rangle$ (open, blue squares).}
\end{figure}
\\\indent Exposing the droplets to ultrashort pulses of 15\,ps duration results in similar energy distributions as for the solid target. Figure~\ref{fig:psDroplet}a shows the distributions for increasing energy density of the laser pulse. The distributions are fit with Eq.~(\ref{MorasModelEq}) and plotted as dashed (black) lines. The agreement between the experimental distributions and the model is good for ion energies below 10\,keV. For high energy densities of the laser \mbox{($>$20\,J/cm$^2$)} Eq.~(\ref{MorasModelEq}) underestimates the amount of ions with energies above 10\,keV. 
Again, the characteristic ion energies are plotted in dependence of the peak laser fluence in Fig.~\ref{fig:psDroplet}b. Below peak fluences of 40\,J/cm$^2$ of the laser the characteristic ion energies increase. At higher peak fluence (100\,J/cm$^2$ ) the fit misses the high-energy tail of the distribution. As a result, the value for $E_0$ obtained from the fit appears to saturate at 1.2\,keV. \added{We find good agreement between the obtained values $\langle E_{\rm exp} \rangle$ and $\langle E_{\rm fit} \rangle$ (see Fig.\,\ref{fig:psDroplet})}.\deleted{below this saturation point (see Fig.\,\ref{fig:psDroplet}).}
\begin{figure}[bt!]
	\centering
	\includegraphics{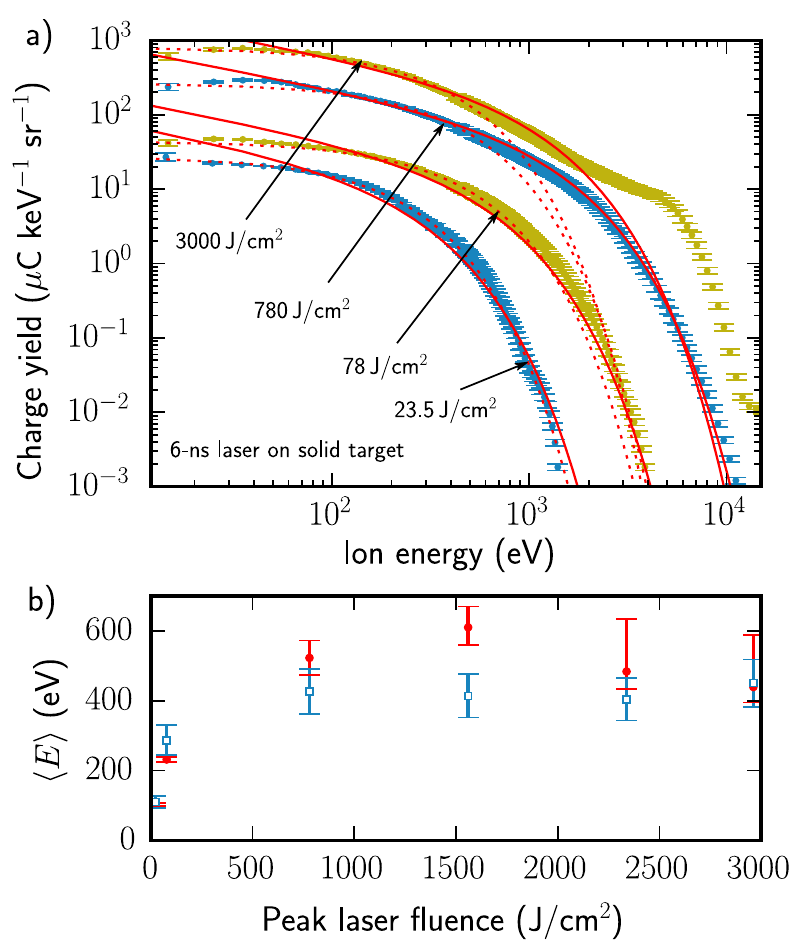}
	\caption{\label{fig:nsonsolid} (a) Charge yield distributions for different peak fluences on the solid target and fits with Eq.~(\ref{MurakamiModelEq}) and $\alpha = 2$ dashed (dark-red) lines, $\alpha = 1$ solid (red) lines. \deleted{The peak fluences are 23.5, 78, 780, 3000\,J/cm$^2$ from bottom to top graph. The pulse duration of the laser pulses is 6\,ns.}
    (b) The values for $\langle E \rangle$ are obtained from the fits with Eq.~(\ref{MurakamiModelEq}) for these distributions. \deleted{Closed (red) triangles correspond to $\tilde{E_{0}}=\langle E_{\rm fit} \rangle$ for $\alpha = 2$ and}Closed (red) circles \added{correspond} to $\tilde{E_{0}}=\langle E_{\rm fit} \rangle$ for $\alpha = 1$, along with $\langle E_{\rm exp} \rangle$ (open, blue squares). Obtained values for $\tilde{E_{0}}$ for $\alpha = 2$ \deleted{ and peak fluences higher than 78\,J/cm$^2$}are omitted.}
\end{figure}
\\\indent The charge distributions change significantly when we use the 6-ns instead of the ps-laser pulses to produce the plasma. Figure~\ref{fig:nsonsolid}a shows the energy distributions derived from the time-of-flight traces of the ions emitted from the solid target at an angle of 2$^{\circ}$. The distributions are measured at peak fluences of the laser pulses ranging from 23.5\,J/cm$^2$ to 3\,kJ/cm$^{2}$. Fitting the distributions with Eq.~(\ref{MurakamiModelEq}) requires to set an appropriate dimensionality parameter $\alpha$. The parameter is determined by the ratio of the typical plasma flow length scale and the size of the laser spot size \cite{Murakami2005}. In our experiments this length scale and laser focus are of similar size and thus the choice of the dimension is not straightforward. We find that setting $\alpha=1$ or 2 gives satisfactory agreement with the obtained data in the following. \added{To determine the actual dimensionality of the expanding plasma, further measurements are required over a range of laser spot sizes with a multi-angle and charge-state-resolved approach.} With the dimensionality parameter set to $\alpha=1$ the energy distributions produced by pulses of laser fluences between 80\,J/cm$^2$ and 1.6\,kJ/cm$^2$ are fit with Eq.~(\ref{MurakamiModelEq}). Examples of the fit with Eq.~(\ref{MurakamiModelEq}) and $\alpha=1$ to the energy distribution are shown as solid (red) lines in Fig.~\ref{fig:nsonsolid}a. For $\alpha=2$ the fit is illustrated by the dashed (red) lines. The energy distributions obtained with laser fluences below 80\,J/cm$^2$ both $\alpha = 1$ and $\alpha = 2$ produce good agreement with Eq.~(\ref{MurakamiModelEq}). The ion energy distribution shows a flat response below 50\,eV, which is better captured by choosing $\alpha = 2$. At peak fluences above 2.4\,kJ/cm$^2$ the energy distributions feature a ``shoulder'' around an energy of 6\,keV that is not reproduced by Eq.~(\ref{MurakamiModelEq}).\\\indent
Figure.~\ref{fig:nsonsolid}b shows the average energies of ions $\left<E_{\rm fit}\right>=\tilde{E}_0/2$ for $\alpha=1$ obtained from fitting the data to Eq.~\ref{MurakamiModelEq} as solid (red) circles. The open (blue) squares show the average energies obtained from the experimental data.\deleted{The obtained values for $\tilde{E_0}$, when setting $\alpha=2$, are shown as open (black) squares. Values for $\tilde{E_0}$ for peak fluences above 80\,J/cm$^2$ are not shown, because the fit misses a large part of the energy distributions at higher ion energies. For these higher peak fluences the fit is performed with $\alpha=1$, and the resulting values for $\tilde{E_{\rm 0}}$ are shown as closed (red) circles.} The characteristic ion energies follow a non-linear trend saturating at a peak fluence of 1.6\,kJ/cm$^2$. Then, at a higher peak fluence the fit becomes inaccurate because of the abundance of ions with energies above 6\,keV. \deleted{The average kinetic energy from Eq.~(\ref{MurakamiModelEq}) is $E_0/2$ for $\alpha=1$.} \added{At the lower fluences, we obtain reasonable agreement between the values $\langle E_{\rm exp} \rangle$ and $\langle E_{\rm fit} \rangle$ (see Fig.\,\ref{fig:nsonsolid}).}
\\\indent Our comparisons between theoretical and measured charge-integrated energy distributions show that over a wide range of peak fluences the results of Mora \cite{Mora2003PlasmaExpansionVacuum} and Murakami \textit{et al.} \cite{Murakami2005} can be employed to characterize ion emission of LPPs. Care should be taken when laser pulses of high peak fluence are used to create LPPs. Under such conditions, the energy distributions exhibit an abundance of charges at high energies. Especially for the 6-ns pulses with energies \mbox{$>$ 2.6\,kJ/cm$^2$} the distribution shows a peak that cannot be reproduced by either of the two model descriptions.
\section{Conclusion}
We present the ion distributions of LPPs for droplet and planar targets for various laser pulse lengths and energies and compare them with the predictions of two results of hydrodynamic models. The charge-integrated energy distributions of ions are well explained by theoretical predictions of Refs.~\cite{Murakami2005,Mora2003PlasmaExpansionVacuum}. The ion energy distributions fit well the energy distributions found by Mora \cite{Mora2003PlasmaExpansionVacuum} when the plasma is produced by laser pulses below 100\,ps. In contrast, laser pulses of 6\,ns duration produced expanding plasmas with ion energy distributions that can be fit by the findings of Murakami \textit{et al.} \cite{Murakami2005}. The essential difference of the expansion of plasma produced either by ultrashort pulses or ns-long pulses lies in the density evolution of the plasma during expansion. Ultrashort pulses produce plasma with an exponentially decaying density. While ultrashort pulses are off when the produced plasma expands, the ns-long pulse continues to heat the disintegrated target during part of its expansion. The density of the plasma generated in this way has a Gaussian shape, and the pressure of the plasma decreases in time. 
The two types of plasma expansions from LPPs may be studied in future to clarify this dynamical behavior in the transition regime by producing plasma with laser pulses between 100\,ps and 6\,ns.\\\indent
Fitting the theoretical findings to the experimental energy distributions provides a characteristic ion energy of the expanding plasma. By performing additional charge-state resolved measurements the actual electron temperature of the plasma, as in Eq. (\ref{eq:eoformula}), may be determined. Charge state resolving ion energy spectrometry not only will enable the determination of the electron temperature, but may point at why the theoretical predictions fail to explain an abundance of high energy ions when the plasma is produced by high-peak-fluence laser pulses.\\\indent
The findings of our work show that relatively simple models are sufficient to explain measured ion energy distributions of the LPPs studied here. The understanding of ion emission of expanding plasmas is an important step to assess optics damage in short-wavelength light sources. 
\section*{Acknowledgement}
\added{We thank M. Basko for fruitful discussions. }
This work has been carried out at the Advanced Research Center for Nanolithography (ARCNL), a public-private partnership of the University of Amsterdam (UvA), the Vrije Universiteit Amsterdam, the Netherlands Organisation for Scientific Research (NWO) and the semiconductor equipment manufacturer ASML.

\section*{References}
\bibliographystyle{ieeetr}
\bibliography{SnIonLiterature} 
\end{document}